\documentclass[onecolumn]{aa}
\usepackage{graphicx}

\begin{document}

\title{Dust in the Edgeworth-Kuiper belt zone}
\author{J. Kla\v{c}ka$^{1}$, L. K\'{o}mar$^{1}$, P. P\'{a}stor$^{1,2}$}
\institute{Faculty of Mathematics,
   Physics and Informatics, Comenius University \\
   Mlynsk\'{a} dolina, 842 48 Bratislava, Slovak Republic \\
   e-mail: klacka@fmph.uniba.sk
   \and
   Tekov Observatory, \\
   Sokolovsk\'{a} 21, 934~01, Levice, Slovak Republic}

\date{}

\abstract{
Orbital evolution of spherical interplanetary dust particles in the Edgeworth-Kuiper belt zone
is treated for semimajor axes 30-50 AU. Besides solar gravity, also solar electromagnetic and 
corpuscular radiation, and, fast interstellar gas flow are important forces influencing 
motion of the particles. The solar electromagnetic radiation is represented by the 
Poynting-Robertson effect and the solar corpuscular radiation corresponds to the solar wind. 
Time-variability of the non-radial solar wind can significantly increase dust lifetime in the zone. 
The average time for the particle stay in the zone is more than 30-times greater than the conventional 
case of constant (time independent) radial solar wind offers, for the particles of tens micrometers in size. 
This holds for the most realistic material properties of the particles: $\bar{Q}'_{pr} =$ 1, where 
$\bar{Q}'_{pr}$ is the dimensionless efficiency factor for electromagnetic radiation pressure. If 
$\bar{Q}'_{pr} =$ 1/2, then the average time of the particle stay in the zone is only 4-times the conventional 
value. The approach used in the paper illustrates the relevance of the solar wind action in comparison with
the Poynting-Robertson effect. The results have an important consequence for our understanding of the
structure of dust distribution in the Edgeworth-Kuiper belt zone, and, also, of dust belts in other stellar
systems.

}

\authorrunning{Kla\v{c}ka et al.}
\titlerunning{Dust in the Edgeworth-Kuiper belt zone}

\maketitle

\section{Introduction}

Orbital evolution of dust particles in the Solar System is investigated, partially in a physical way, 
since the time of Poynting (1903). Besides action of solar gravity and solar electromagnetic 
radiation, in the form of the Poynting-Robertson effect (Poynting 1903, Robertson 1937,
Kla\v{c}ka 2004, 2008a, 2008b, Kla\v{c}ka et al. 2009a), also the effect
of solar corpuscular radiation is taken into account (Whipple 1955, 1967, Dohnanyi 1978 and others).
It is generally considered that the effect of solar wind is similar to the effect of solar electromagnetic radiation
and the solar wind is about (0.2-0.3)-times less important than the Poynting-Robertson effect
(Whipple 1955, 1967, Dohnanyi 1978, and, e.g., Mukai and Yamamoto 1982, Jackson and Zook 1989, 
Leinert and Gr\"{u}n 1990, Dermott et al. 1994, Gustafson 1994, Reach et al. 1995,
Dermott et al. 2001, Minato et al. 2004, Abe 2009, Mann 2009).
However, this conventional idea is in variance with reality, in general 
(Kla\v{c}ka and Saniga 1993, Kla\v{c}ka 1994, Kocifaj and Kla\v{c}ka 2008, 
Kla\v{c}ka et al. 2009b, Kla\v{c}ka et al. 2009c, P\'{a}stor et al. 2010). 
It turns out that orientation on the correct physical treating of the solar wind effect
(Kla\v{c}ka et al. 2009c) can significantly improve our knowledge of the action of the solar wind 
on the orbital evolution of dust grains in the Solar System. This paper is based on the theory presented by
Kla\v{c}ka et al. (2009c). 

The aim of this paper is to apply solar wind action on orbital evolution of dust particle moving in the zone 
of the Edgeworth-Kuiper belt, which can be defined by semi-major axes 30-50 AU, approximately. 
We are interested in the time of the particle stay in the zone. More physical model of the solar wind action 
(Kla\v{c}ka et al. 2009c) will be compared with the conventional access. 
The particles of several micrometers to tens of micrometers are considered, so the Lorentz force
(interplanetary magnetic field) and collisions among particles are neglected (Dohnanyi 1978, 
Leinert and Gr\"{u}n 1990, Dermott et al. 2001 and Gr\"{u}n et al. 1985). Gravity of the Sun,
Poynting-Robertson effect, action of the solar wind and fast interstellar gas flow are forces which influence
motion of the particles. Simultaneous action of all these effects can shed a light on the relevance of the
solar wind on the orbital evolution of dust particles. The importance of the solar wind action
and the P-R effect can be compared. The results can help not only in better understanding of 
dust evolution beyond planets in the Solar System, but also in understanding of dust belts in other stellar 
systems (see, e.g., Marley 2008).

Section 2 presents relevant equation of motion. Sec. 3 offers the most important results of numerical solution 
of the equation of motion. The section compares the results for the conventional time-independent 
radial solar wind with those obtained for more real solar wind model. Sec. 4 presents a short discussion
on current situation in modeling of dust evolution in the Solar System.

\section{Equation of motion}

Dealing with orbital evolution of dust particles in space beyond planets,
gravity of the Sun and nongravitational forces acting on the particles play 
the most relevant role. Equation of motion of the dust grain under the action of
solar gravity, solar electromagnetic and corpuscular radiation,
fast interstellar gas flow, and gravity of other bodies, is
\begin{eqnarray}\label{eqm}
\frac{d ^{2} \vec{r}}{dt^{2}} &=& -~ \frac{G~M_{\odot}}{r^{3}} ~\vec{r} 
\nonumber \\
& & +~ \beta ~ \frac{G~M_{\odot}}{r^{2}} ~
      \left \{ \left ( 1~-~
      \frac{\vec{v} \cdot \vec{e}_{R}}{c} \right ) \vec{e}_{R}
      ~-~ \frac{\vec{v}}{c} \right \} 
\nonumber \\
& & +~ \beta ~\frac{\eta}{\bar{Q}'_{pr}} \frac{G~M_{\odot}}{r^{2}} ~
\frac{1}{c~u} ~ | \vec{u} ~-~ \vec{v} | ~ \left ( \vec{u} ~-~ \vec{v} \right )
\nonumber \\
& & +~ c_{D} ~\gamma_{H} ~ | \vec{v}_{H} ~-~ \vec{v} | ~ \left ( \vec{v}_{H} ~-~ \vec{v} \right ) ~,
\nonumber \\
& & -~ \sum_{i = 1} ^{N} G~M_{i}~ \left ( \frac{\vec{r} ~-~\vec{r}_{i}}{| \vec{r} ~-~\vec{r}_{i} |^{3}} 
~+~ \frac{\vec{r}_{i}}{| \vec{r}_{i} |^{3}} \right ) ~,
\end{eqnarray}
where $\vec{r}$ is the grain's position vector with respect to the Sun, velocity vector
$\vec{v}$ $=$ $d\vec{r}/dt$, $G$ is the gravitational constant, $M_{\odot}$ is mass
of the Sun, $r$ $=$ $| \vec{r} |$. 

The first term on the right-hand side of Eq. (\ref{eqm}) corresponds to gravitational acceleration of the Sun.

The second term on the right-hand side of Eq. (\ref{eqm}) is the action of the solar electromagnetic radiation
on a spherical particle. It is represented by the Poynting-Robertson effect, to the first order in $\vec{v}/c$, 
where $c$ is the speed of light in vacuum. The unit radial vector $\vec{e}_{R}$ is defined by the ratio
$\vec{r}/ r$, $\vec{e}_{R}$ $\equiv$ $\vec{r}/ r$. 
The quantity $\beta$ is a dimensionless parameter defined as the ratio of
radial component of the radiation force and the gravitational force between the Sun and 
the particle with zero velocity. For homogeneous spherical particle 
\begin{eqnarray}\label{PR}
\beta &=& \frac{L_{\odot} ~A' ~\bar{Q}'_{pr}}{4 ~\pi ~c~ m ~G~M_{\odot}} ~,
\nonumber \\ 
\beta &=& 5.76 \times 10^{2}
\frac{\bar{Q}'_{pr}}{R [\mu \mbox{m}]~\rho [\mbox{kg/m}^{3}]}~,
\end{eqnarray}
$L_{\odot}$ $=$ 3.84 $\times$ 10$^{26}$ W is luminosity of the Sun (Bahcall 2002), 
$\bar{Q}'_{pr}$ is a dimensionless efficiency factor for electromagnetic radiation pressure,
$A'$ $=$ $\pi$ $R^{2}$ is geometric cross section of the homogeneous spherical particle of
radius $R$ and mass density $\rho$, $m$ $=$ (4/3) $\pi$ $\rho$ $R^{3}$. 
We assume that $\beta$ $=$ $const$:
neither optical properties nor mass of the dust particle change. We have
\begin{eqnarray}\label{cpr}
\bar{Q}'_{pr} &=& \bar{C}'_{pr} / A' ~,
\nonumber \\
\bar{C}'_{pr} &=& \frac{\int_{0}^{\infty} I \left ( \lambda \right ) ~
		     C'_{pr} \left ( \lambda \right ) ~ d \lambda}{
		     \int_{0}^{\infty} I \left ( \lambda \right ) ~ d \lambda} ~,
\end{eqnarray}
where $\bar{C}'_{pr}$ is spectrally averaged cross section of radiation pressure
and $I ( \lambda )$ is the flux of monochromatic radiation energy as a function of wavelength. We treat 
the homogeneous spherical particle when non-radial components of the cross section vector of radiation pressure 
are zero -- this is known as the Poynting-Robertson effect (Poynting 1903, Robertson 1937,
Kla\v{c}ka 2008b, Kla\v{c}ka et al. 2009a).
Thus, the solar electromagnetic radiation is represented by the Poynting-Robertson effect in our paper.
The quantity $\bar{Q}'_{pr}$ can be calculated according to Mie (Mie 1908; see also van de Hulst 1981, 
Bohren and Huffman 1983), or, a more general theory can be found in Kla\v{c}ka and Kocifaj (2007).
The more general theory includes also electric charge of the spherical particle. This paper considers that
surface electric potential of the particle is zero. This is a good approximation to reality for the particle
sizes of tens of micrometers.

The third term on the right-hand side of Eq. (\ref{eqm}) is the action of the solar corpuscular radiation,
solar wind. We take into account the equation of motion derived by Kla\v{c}ka et al. (2009c), but
we neglect decrease of dust particle's mass. While $\eta$ is conventionally a constant of the value 0.22 
(Whipple 1967, Dohnanyi 1978), or, 0.30 (e.g., Gustafson 1994, Abe 2009), or, 1/3 
(e.g., Kla\v{c}ka et al. 2009b) and the solar wind speed values of 350 km/s (Dohnanyi 1978)
or 400 km/s (Gustafson 1994) are used, we take into account more precise solar physics observational data 
(Svalgaard 1977 -- chapter 13, Hundhausen 1997, p. 92, Bruno et al. 2003). The final results are given in 
Eqs. (56)-(57) by Kla\v{c}ka et al. (2009c). Thus, we have
\begin{eqnarray}\label{sw}
u ( r, t ) &=& u_{0} \left \{ 1 - \delta \cos \left [ 2 \pi ~ \frac{t - r / \left ( 2 u \left ( r, t \right ) 
\right ) - t (max)}{T} \right ] \right \} ~, 
\nonumber \\
\eta (r, t) &=& \eta_{0} [u(r, t) / u_{0}]^{2} ~,
\nonumber \\
\eta_{0} &=& 0.38 ~,
\nonumber \\
\delta &=& 0.15 ~,
\nonumber \\
u_{0} &=& 450~ \mbox{km/s} ~,
\nonumber \\
T &=& 11.1~ \mbox{years} ~.
\end{eqnarray} 
$T$ is the average value of the solar cycle period (see, e. g., Foukal 2004, p. 366),
$t (max)$ is the moment/instant of the solar cycle maximum.
The retarded time $t_{retard}$ $\equiv$ $r / [ 2 u ( r, t ) ]$ is of the order 
$r / u_{0}$. We may put $t_{retard}$ $=$ $r$ $/$ (2$u_{0}$) as a very good approximation
and this value is used in numerical calculations. Moreover, we use the decomposition of the 
velocity vector $\vec{v}$ $=$ $v_{R} \vec{e}_{R}$ $+$ $v_{T} \vec{e}_{T}$,
and, the used solar wind velocity vector $\vec{u}$ is 
\begin{eqnarray}\label{bruno}
\vec{u} &=& ( \gamma_{R} ~\vec{e}_{R} ~+~ \gamma_{T} ~\vec{e}_{T} ) ~ u (r ,t) ~,
\nonumber \\
\gamma_{T} &=& 0.05 ~, 
\nonumber \\
\gamma_{R} &=& \sqrt{1 ~-~ \gamma_{T} ^{2}} \doteq 1 ~. 
\end{eqnarray}
The approximation of a constant flux for the radial solar wind
is described by the conditions $\eta$ $=$ $\eta_{0}$ $=$ 0.38, $\gamma_{T}$ $=$ 0.

The fourth term on the right-hand side of Eq. (\ref{eqm}) corresponds to the action of the 
fast interstellar gas flow (Fahr 1996, Lee et al. 2009, P\'{a}stor et al. 2010).
The drag coefficient $c_{D}$ can be approximated by the value 2.6 (Baines et al. 1965,
Scherer 2000, Kla\v{c}ka et al. 2009b, P\'{a}stor et al. 2010). More exact result is given by the formula
\begin{eqnarray}\label{cdrag}
c_{D} &=& \frac{\xi}{M} ~ \left ( C_{1} ~+ C_{2} \right ) ~,
\nonumber \\
C_{1} &=& \left ( 1 ~+~ \frac{1}{2 M^{2}} \right ) ~ \frac{\exp \left ( -~M^{2} \right )}{\sqrt{\pi}} ~,
\nonumber \\
C_{2} &=& M ~ \left ( 1 ~+~ \frac{1}{M^{2}} ~-~ \frac{1}{4 M^{4}} \right ) ~ erf (M) ~, 
\nonumber \\
M &=& | \vec{v}_{H} ~-~ \vec{v} | / \sqrt{2 k_{B} T_{H} / m_{H}} ~, 
\end{eqnarray}
where $\xi$ is the adsorption or sticking factor (1 $\le$ $\xi$ $\le$ 2) which describes the relevant type
of collision (i.e. specular $\xi$ $=$ 1, diffuse $\xi$ $=$ 1/2, or adsorption $\xi$ $>$ 1, Baines et al. 1965;
unfortunately, Scherer (Scherer 2000) presents incorrect value for the diffuse type of collisions, $\xi$ $=$ 1; 
standardly used value is $\xi$ $=$ 2 - Scherer 2000). We use the value $\xi$ $=$ 2 in Eq. (\ref{cdrag}). 
$M$ is the Mach number, $erf (M)$ is the error function, $erf (M)$ $=$ ($2/\sqrt{\pi}$) $\int_{0}^{M} \exp (-~t^{2} ) ~dt$,
$\vec{v}_{H}$ is the velocity vector of the interstellar hydrogen atoms
($| \vec{v}_{H} |$ $\approx$ 25 km/s), $k_{B}$ is the Boltzmann constant,
$m_{H}$ is mass of the hydrogen atom, $T_{H}$ is the temperature of the interstellar gas,
$T_{H}$ $\approx$ 8 $\times$ 10$^{3}$ K (Lallement 1990, 1996).
The collisional parameter $\gamma_{H}$ is $\gamma_{H}$ $=$ $n_{H} m_{H}$ $A'$ / $m$, 
where the concentration of the neutral hydrogen atoms is $n_{H}$ $=$ 0.05 cm$^{-3}$ in the outer heliosphere
$r$ $\in$ (30 AU, 80 AU) (Fahr 1996). 

Finally, the fifth term on the right-hand side of Eq. (\ref{eqm}) corresponds to the gravitational accelerations 
from other bodies, e.g., planets, of masses $M_{i}$, with heliocentric positional vectors $\vec{r}_{i}$
($i$ $=$ 1 to $N$). 

Initial conditions must be added to the equation of motion. If the particle is ejected from
a parent body of known orbital elements, then the particle's initial orbital elements have to be calculated 
from Eqs. (60)-(61) in Kla\v{c}ka (2004).

\section{Numerical results}

This section concentrates on orbital evolution of interplanetary dust particle under the action of 
solar electromagnetic radiation, solar corpuscular radiation when solar wind erosion is neglected, and,
fast interstellar gas flow. The results based on Eqs. (\ref{eqm})-(\ref{cdrag}) are compared with the conventional 
approach when only radial solar wind with constant $\eta$ is taken into account.

Uniform distributions in semi-major axis, $a$ $\in$ (30 AU, 50 AU), and eccentricity,
$e$ $\in$ (0, 0.5), are taken as the initial conditions for particles distributions in the two orbital elements.
Planar problem is considered. 

\begin{table}[h]
\centering
\begin{tabular}{|c|c|c|c|c|}
\hline
Forces & \multicolumn{2}{c|} {$\beta = 0.1$} &
\multicolumn{2}{c|} {$\beta = 0.01$} \\
\cline{2-5}
& $\bar{Q}'_{pr} = 1$ & $\bar{Q}'_{pr} = 1/2$ & $\bar{Q}'_{pr} = 1$ & $\bar{Q}'_{pr} = 1/2$ \\
\hline
P-R + const. SW (R) + IG & 9.3E+05 & 6.3E+05 & 9.2E+06 & 6.2E+06 \\
\hline
P-R + var. SW (R) + IG & 1.5E+06 & 1.1E+06 & 1.5E+07 & 1.1E+07 \\
\hline
P-R + const. SW (NR) + IG & 2.7E+06 & 4.3E+06 & 2.4E+07 & 3.4E+07\\
\hline
P-R + var. SW (NR) + IG & 1.9E+07 & 2.3E+06 & 3.2E+08 & 2.7E+07 \\
\hline
\end{tabular}
\caption{Mean times (years) of spiralling of dust particles within
the Edgeworth-Kuiper belt: $a$ $\in$ (30 AU, 50 AU). Two values of $\beta$
and $\bar{Q}'_{pr}$ are used. The Poynting-Robertson effect, the effect of 
constant ($\eta$ $=$ 1/3, Kla\v{c}ka et al. 2009b) and time-variable solar wind 
(radial and non-radial, $\gamma_{T}$ $=$ 0 and 0.05) 
and the effect of neutral interstellar gas flow are considered.}
\label{tab:1}
\end{table}

Relevant results are summarized in Table 1, where average lifetime of dust particle
in the considered zone is presented. Various properties of the particle and various 
combinations of non-gravitational forces (besides gravity of the Sun) are considered.
As for the dimensionless efficiency factor for electromagnetic radiation pressure,
the value $\bar{Q}'_{pr} =$ 1 is a good approximation to reality. The conventional
approach corresponds to the first line. However, the physical approach
is represented by the fourth line. 

If one considers a particle of tens of micrometers in diameter ($\beta$ $=$ 0.01), then Table 1 immediately yields 
that the average lifetime of the particle in the considered zone is 35-times 
(3.2 $\times 10^{8}$ years / 9.2 $\times 10^{6}$ years) greater than the conventional lifetime, 
for the case $\bar{Q}'_{pr} =$ 1. Similar result is obtained for 10-times larger particle when $\beta$ $=$ 0.1.

Table 1 offers other interesting results. If one would consider $\bar{Q}'_{pr} =$ 1/2, then the average lifetime 
would be only 4-times (e.g., 2.7 $\times 10^{7}$ years / 6.2 $\times 10^{6}$ years for $\beta$ $=$ 0.01) 
greater than the conventional lifetime. However, the case $\bar{Q}'_{pr} =$ 1/2 corresponds to optical properties 
far from reality. Also time-non-variable (time independent / constant) non-radial solar wind (see the third line)
does not produce such stability as the physical case of time-variable non-radial solar wind (see the fourth line).

Relative values of the lifetimes, presented in Table 1, are relevant. The reason is that the lifetime 3.2 $\times 10^{8}$ 
years is too large as for time action of the insterstellar gas cloud on the Solar System: both as for the position of the 
cloud with respect to the Solar System and as for the dimension ($v_{H}$ times lifetime) of the cloud.
In any case, real structure of dust disk does not correspond to consideration of time-independent
radial solar wind in computational modeling. 

We have the following important summarization of Table 1. If one considers physical approach to solar wind
effect, based on the correct physical theory respecting solar physics observational data (Kla\v{c}ka et al. 2009c), 
then the lifetime of a dust particle in the region beyond planets is more than 30-times greater than the conventional 
approach offers. However, also correct optical properties of the dust particle have to be taken into account 
($\bar{Q}'_{pr} \doteq$ 1). All these facts have an important consequence for the structure of dust distribution in the 
Edgeworth-Kuiper belt zone. Synergy of several physical approaches can help in much better understanding of reality 
than a simple heuristic approach can offer.

 \section{Discussion}

Previous section presents relevant new results. However, the results are obtained as a solution of 
 Eq. (\ref{eqm}) without gravity of planets. Really, if one would like to obtain complete realistic 
idea about evolution of dust particles beyond (or, mainly, inside) the zone of planets, then one should take 
into account also gravitational perturbations of planets, too. However, the aim of the paper is to point out on 
the importance of correct treatment of non-gravitational forces. In reality, besides inclusion of planets, also
non-planar case has to be considered. Not only as for gravitational, but also as for 
non-gravitational accelerations. The real 3-dimensional problem for the Poynting-Robertson effect and
the effect of the neutral interstellar gas flow are correctly understood (P\'{a}stor et al. 2010).
However, this cannot be said about the solar wind effect. It will require much effort for future work.
Before this is correctly done, any modeling of the dust evolution beyond planets is only 
a kind of an exercise. Thus, our results presented in the previous section have to be understood as
an important step toward the physical understanding of reality.

 \section{Conclusion}

The paper presents an application of the solar wind action on orbital evolution of dust particles
in the Solar System. 

Conventional idea is that the solar wind effect is similar to the Poynting-Robertson (P-R) effect 
and the P-R effect is more relevant than the solar wind effect.
Our results show that the conventional idea is far from reality. More physical results yield that 
dust particle can stay in the Edgeworth-Kuiper belt zone more than 30-times longer than it would be 
obtained by the conventional consideration of the non-variable (constant in time) radial solar 
wind, if the solar gravity, the Poynting-Robertson effect, the solar wind effect and
fast interstellar gas flow are considered accelerations influencing motion of the particles. 
Thus, one has to take into account more physical approach than 
the conventional one, when dealing with orbital evolution of dust particles. 

The results show an importance of the correct treating of non-gravitational effects acting on dust particles.
Although gravity of a planet is of non-negligible importance, it cannot change some kind of 
stabilization of dust beyond the zone of planets caused by the non-gravitational effects. 
The greater heliocentric distance of the particles, the less important is the effect of the planet.
[Inclusion of the planet Neptune yields for the mean lifetime in the Edgeworth-Kuiper belt zone
($\beta =$ 0.01 and $\bar{Q}'_{pr} =$ 1): 
3.1 $\times$ 10$^{7}$ years for the simplest approach (P-R effect +
radial time-independent solar wind + interstellar gas + gravity of the Sun and Neptune), and,
7.3 $\times$ 10$^{7}$ years for the physical approach (P-R effect +
non-radial time-variable solar wind + interstellar gas + gravity of the Sun and Neptune), if only
dust particles not collided with Neptune and not ejected from the Solar System (due to the gravity of Neptune)
are taken into account.]

The found results are important also for understanding of dust disks/belts evolution in stellar systems.

\section*{Acknowledgement}
This work was partially supported by the Scientific Grant Agency VEGA, Slovakia,
grant No. 2/0016/09.

\end{document}